\documentclass[letterpaper, onecolumn, 9pt]{article}

\usepackage[english]{babel}
\usepackage[utf8]{inputenc}
\usepackage[T1]{fontenc}
\usepackage[left=1in, right=1in, top=0.5in, footskip=0.5in]{geometry}
\usepackage{graphicx}
\usepackage[space]{grffile}
\usepackage{epstopdf}
\usepackage{wrapfig}
\usepackage{multirow}
\usepackage{listings}
\usepackage{setspace}
\usepackage{indentfirst}
\usepackage{rotating}
\usepackage{hyperref}
\hypersetup{
    colorlinks=true, 
    linktoc=all,     
    linkcolor=blue,  
    allcolors=blue,
}
\usepackage{nameref}
\usepackage{mathtools}
\usepackage{amsmath,amsthm}
\usepackage[makeroom]{cancel}
\usepackage{amsfonts}
\usepackage{amssymb}
\usepackage[font=small,labelfont=bf,format=plain,justification=justified]{caption}
\usepackage{floatrow}
\usepackage{subcaption}
\usepackage{array}
\newcolumntype{H}{>{\setbox0=\hbox\bgroup}c<{\egroup}@{}}
\usepackage{pstricks}
\usepackage{epic}
\usepackage{enumerate}
\usepackage{booktabs}
\usepackage{longtable}
\usepackage[square,numbers,comma,sort&compress]{natbib}
\usepackage{authblk}
\usepackage{ulem} 
\usepackage{siunitx}
\usepackage{tabularx} 
\usepackage{booktabs}

\definecolor{forestgreen}{rgb}{0.33,0.61,0.34}

\usepackage[english]{babel}

\date{}

\usepackage{amsmath}
\usepackage{graphicx}

\title{Shaping New Norms for AI}
\author[1,2]{Andrea Baronchelli}
\affil[1]{City University of London (UK)}
\affil[2]{The Alan Turing Institute, London (UK)}

\begin{document}
\maketitle

\begin{abstract}
As Artificial Intelligence (AI) becomes increasingly integrated into our lives, the need for new norms is urgent. However, AI evolves at a much faster pace than the characteristic time of norm formation, posing an unprecedented challenge to our societies. This paper examines possible criticalities of the processes of norm formation surrounding AI. Thus, it focuses on how new norms can be established, rather than on what these norms should be. It distinguishes different scenarios based on the centralisation or decentralisation of the norm formation process, analysing the cases where new norms are shaped by formal authorities, informal institutions, or emerge spontaneously in a bottom-up fashion. On the latter point, the paper reports a conversation with ChatGPT in which the LLM discusses some of the emerging norms it has observed. Far from seeking exhaustiveness, this article aims to offer readers interpretive tools to understand society's response to the growing pervasiveness of AI. An outlook on how AI could influence the formation of future social norms emphasises the importance for open societies to anchor their formal deliberation process in an open, inclusive, and transparent public discourse.
\end{abstract}

\vspace{0.8cm}

\begin{quote}
   {\textit{What is the social norm for using it? What are the legal norms?} \\
  {Jensen Huang, CEO of Nvidia, January 2023 \cite{bloomberg2023chatgpt}.}}
\end{quote}

\begin{quote}
    {\textit{When it comes to human interaction with LLMs like myself, \\ there are evolving social norms that have started to emerge.
} \\
  {ChatGPT. Conversation with the author, May 2023.}}
\end{quote}

\vspace{0.5cm}

\section{Introduction}

It is likely that 2023 will be remembered as the year of Artificial Intelligence (AI). ChatGPT \cite{openai-chatgpt} was the fastest internet service to reach 100million users until now (May 2023) \cite{reuters2023chatgpt} and the technology of Large Language Models (LLMs) at its core is a fundamental element of sister apps for images such as Dall-e2, Midjourney and many others. One of the most fascinating aspects of LLMs is that they exhibit unpredicted emergent features. While the media excitedly reported how AI art generator have developed their own taste \cite{economist2023artificialintelligence} or chatbots are able to pass school level exams in a growing number of disciplines \cite{cnn2023chatgpt}, only in 2023 it was released that, for the past two years, GPT models had consistently improved its performance in tests designed to measure theory of mind in children \cite{kosinski2023theory}. 

For anyone familiar with complexity science, observing emergent properties in a complex system made of billions of artificial neurons is perhaps not surprising, but the growth in human-, or even superhuman-, like capabilities has attracted huge attention from the media and the public, sparking a hectic debate between the technology apocalyptic and integrated \cite{eco_apocalypse_postponed}. While it is clear that AI could bring us spectacular benefits, from better medical diagnosing to drug discovering, the risks have so far catalysed most of the public attention. Perils associated to narrow AI include increasing opportunities for manipulation of people, enhancing and dehumanising weapons, and rendering human labour increasingly obsolescent \cite{Federspiele010435}. On the other hand, self-improving "artificial general intelligence" (AGI) could pose an existential threat to humanity itself.

Despite the prevailing uncertainty, there is consensus on at least two points. First, AI is going to have a significant impact on our life. Second, society is not prepared to deal with the technology \cite{statement}. New rules are needed to help the transition towards a world where humans and machine coexist to the benefit of the former, if not of both parties. The call for action comes from some of the leading figures of the AI revolution. In an official note, OpenAI, the company behind chatGPT, recognise an existential risk associated to Artificial General Intelligence and advocate for a strong public oversight over the governance of the "most powerful systems" \cite{openai}. Similarly, Alphabet CEO Sundar Pichai wrote that AI is "too important not to regulate" \cite{ft_sundar}. Earlier in the year, NVIDIA CEO Jensen Huang said that AI will create tools "that require legal regulation and social norms that have yet to be worked out" \cite{bloomberg2023chatgpt}, highlighting how social norms can be a solution to solve large-scale problems \cite{nyborg2016social,andrighetto2022research}.

How to shape these new norms? In the rest of the paper, I will explore some aspects of the onset of new norms for AI from a complex system perspective. Following recent results from the complex systems approach to the emergence of linguistic norms \cite{amato2018dynamics}, I will distinguish on whether the new norm is imposed by a formal authority (e.g., in the form of a regulation or law), by an informal authority (e.g., through social influence) or rather emerges spontaneously in a bottom-up fashion. In all cases, I will not discuss what norms would be desirable or not, but rather focus on some potentially critical aspects of the norm formation process. Before proceeding, two caveats are in order. The first is that I will use the term "norm" in a broad sense, encompassing both regulations and conventions, although of course this is valid only as a first-order approximation \cite{przepiorka2022norms} and neglects the distinction between personal, descriptive, injunctive and other types of social norms. The second caveat is that, regarding AI's novelty, the situation we are facing falls somewhere between norm change, when a new norm replaces and old one, and norm emergence, the case in which a new norm is established in a context where there were none. Although these cases share several similarities, both resulting in the establishment of a new norm, they also present theoretical differences involving the mechanisms of spontaneous emergence and tipping point, respectively, that I will not explore here \cite{baronchelli2018emergence}.

\section*{Formal institutions}

Formal institutions, such as governments, are created with the aim to govern human behaviour, and are endowed with the means to produce regulations - or laws - and enforce them\footnote{Different definitions are possible for formal, as well as informal, institutions. Here, and in the remainder of the paper, we stick to a coarse distinction which has proven to be valid empirically \cite{amato2018dynamics}.}. Top-down formal regulation of AI is what the industry has been vocal about for the past months, amidst claim of existential threats to humanity and legal uncertainty on liabilities caused by AI \cite{CERKA2015376,glaubitz2021should}. 
There is agreement that regulation is needed for a safe adoption of AI in established sectors such as health and finance, to guarantee that the use of AI does not conflict with human rights, existing laws and other ethic considerations. Auditing, i.e., conducting independent and systematic evaluations of an entity's actions or properties and communicating the results to stakeholders, is a promising pathway in this sense \cite{audit}. At the moment of writing, the European Union is at the forefront of the regulation race \cite{cnbc2023euaiact}. On the other hand, it is less clear how to eliminate the \textit{existential threat} that AI poses to humanity, an aspect that has gained most of the public attention. There are at least three main issues making the regulation of super-intelligent, or also current, AI hard to implement in an effective way.

\textbf{Unknown unknowns.} The first problem is that it is difficult to imagine what AI will be able to do, even just in the near future. Technology is often faster than regulation, as cryptocurrencies have shown us for over a decade now, and AI seems poised to be one of the fastest evolving technologies humanity has ever seen. History is full of examples of bad regulation stemming from hurry. For example, in the United Kingdom, the Locomotive Act 1865, popularly known as the Red Flag Act, mandated speed limits for self-propelled vehicles -- i.e., cars -- of 4 mph (6 Km/h) on country roads and 2 mph (3 Km/h) in cities \cite{agnew2020steam}. Vehicles required a three-person crew consisting of a driver, a stoker, and a flagman who, carrying a Red flag, would walk 60 yards (55 m) ahead of the vehicle. This ensured the vehicle's speed aligned with the flagman's pace, enabling them to warn horse carriages and facilitate the vehicle's stoppage until they safely passed \cite{agnew2020steam}. As much as this sounds funny, such an act remained in place for 31 years, until it was repealed in 1896. In 1884, the US state of Vermont proposed a similar act which also lasted until 1896. In the case of AI, the risk that regulation grossly misses the target, being either too strict or too loose, is high.

\textbf{Limited regulatory control.} Analogies are often drawn between the need to regulate AI and the existing approaches to limit the risk of nuclear war or airplane accidents \cite{openai}. There is, however, a substantial difference between the AI industry and many of the examples we are familiar with.
 AI is relatively easier to implement and deploy, leaving virtually no traces. It is mainly a private endeavour, in contrast to  government-run sectors like nuclear power. Additionally, thanks to the internet, AI is not bound by geography. Finally, while large-scale LLM models may require substantial resources, more focused alternatives can be way less demanding. Hence, while formal institutions such as governments can perhaps impose regulations concerning the narrow, "official", use of AI, for example on the workplace, schools and hospitals, it is hard to see how the rule of law might limit malicious actors willing to use AI to commit crimes. For example, while regulation can help forcing online platforms to meet good security standards, there is little doubt that there will be efforts to use AI to organise misinformation campaigns and spread deep fakes. One could object that this is always the case, and that today an airplane can be used to bring destruction. While valid, this objection violates the assumption that AI poses an existential threat to humanity and neglects that it is the first smartphone-compatible technology to do so. Thus, for example, invites to halt the development of self-improving artificial general intelligence \cite{Federspiele010435} are difficult to implement, even where there were a will to do so, due to the opacity of the research taking pace behind closed doors.

\textbf{Quis custodiet ipsos custodes?} Who will guard the guards themselves? The old dilemma of every governance system seems poised to be critical in the case of AI \cite{audit}. While the technology is fragmented and open source versions of AI exist, the bottlenecks of talent and energy costs have concentrated cutting edge research in the hands of few private companies, creating power imbalances. Thus, the statement in favour of public oversight of AI that we mentioned above, posted by OpenAI on May 23, 2023, was followed two days later by the threat that OpenAI might leave the Europe if the latter imposed too much regulation \cite{openai_eu}. Similarly, the above-mentioned piece by Google's CEO Sundar Pichai does not just state that AI is "too important not to regulate", but adds that it is also "too important not to regulate \textit{well}" (italics my own). Who should be the judge for this "well" is not clarified, and the suspects that Google itself may want to retain a voice in the legislative debate is strong. Such a position is far from absurd, given the complexity of the subject, but it highlights obvious conflicts of interest. In this perspective, a possibly illuminating example of the difficult dialogue between regulation and digital technology is the announcement, dated May 27, 2023, that Twitter - the social media platform - will leave EU voluntary Code of Practice on fighting misinformation \cite{EU_twitter}: Regulation has been maintained on a voluntary basis not to scare oligopolists, with the risk that the latter adhere to it just until when they don't.

\section{Informal Institutions}
Another main driver of norm change is represented by informal institutions \cite{amato2018dynamics}. These organisations can not enforce the adoption of a new norm, or can do so only within their non-exclusive remit, yet are influential in proposing new codes of behaviour. For example, in most of today's Western countries, religious leaders can prescript codes of conduct, but have no legal permission to police behaviour (we do not enter here on the psychological aspects that may of course play a major role in such cases). Similarly, a university may produce a code of conduct that includes sanctions for violators, but it cannot prevent students from joining another university with a different set of rules. With their local regulations, typically informal institutions either make up for the lack of formal regulation, as it is currently the case for AI, or detail it further in order to make it fitter to their own environment. Two interesting sectors for an academic audience, among many, where informal institutions have been faced with the urgent need of new norms for AI are scientific publishing and education. 

Journal editors, researchers and publishers started debating about LLMs in the published literature early on: Should ChatGPT be listed as author? Should authors disclose its use? A number of influential publishers, such as Springer, Science and JAMA, concluded right away that ChatGPT can not be listed as an author \cite{ChatGPTAuthor,10.1001/jama.2023.1344,doi:10.1126/science.adg7879}. Motivations range from the practical level "attribution of authorship carries with it accountability for the work, and AI tools cannot take such responsibility" (Nature journals, \cite{Stokel-Walker2023}) to more philosophical considerations that "the [scientific] product must come from—and be expressed by—the wonderful computer in our heads" (Science, \cite{doi:10.1126/science.adg7879}). In a blatant act of anglophone-centric deliberation, which ignores the liberating potential of LLMs as text editors for non-native English speakers, proposed policies regarding the use of LLMs range from banning any text written by AI to ask authors to disclose its usage in the paper, when no existing rule mandates to disclose human proof-editing of a published article \cite{AIPolicies}. Of course, the main limitation of such policies is that compliance is hard to verify.

Education is another heated battleground on how to regulate AI. Several institutions in countries ranging from France to the United States and, most recently, to India and Italy, have moved to outlaw ChatGPT completely, raising concerns that they may limit student access to a formidable learning tool \cite{AIHigherEd,baidoo2023education}. But the cat is out of the box, and AI assistants are doomed to remain a huge temptation for students facing essay-based homework \cite{cano2023chatgpt}. Countries where oral examinations are the norm, such as France or Italy, find themselves in a stronger position. Given that oral exams represent a natural antidote to the LLM doping for students, the question on whether they will face wider adoption worldwide is natural. The fact that different universities may in principle adopt different approaches could represent an ideal experimental setup to evolutionarily determine what the best approach is, yet contagion effects are strong, and not all universities are equally influential. For now, it is hard to imagine that outright bans are a definitive solution. While conservative approaches have been preferred in the beginning of the revolution, more nuanced approaches are likely to emerge soon.

These examples reveal some critical aspects of regulation by informal institutions. Firstly, informal institutions need to act immediately. As soon as a tool such ChatGPT is released, students will use it and Higher Education Institutions are forced to take a position. Secondly, and partly as a consequence of the previous point, informal institutions most often aim to produce regulation tailored on the status-quo of the technology, with no intent to provide a framework able to accomodate future developments. While this may not be a problem for the single institution, which can update their regulation at any point, the consequences can be undesirable for the stakeholders. It is easy to imagine how a student will be told to follow very different, if not contradictory, rules concerning AI during their career. Thirdly, complex contagion effects between institutions operating in the same sector, such as publishing or education, may be strong. As institutions watch one another, standards risk originating by either the first mover or the most prestigious, by some conventional metric, institutions, rather than be the result of an evolutionary process selecting the most desirable norms. In other words, and sticking to the education example, only a tiny fraction of the huge potential universities have in terms of expertise and strategic thinking would be used, as most institutions might just copycat the deliberations made elsewhere.

\section{Spontaneous norms} 
The third main process of norm creation - and change - involves spontaneous emergence. Universally accepted norms are the unintended consequence of individuals’ efforts to coordinate locally with one another~\cite{wittgenstein,steels1995self,baronchelli2006sharp,sugden1989spontaneous,centola2015spontaneous}. Similar to other emergent phenomena observed in complex systems, global coordination in this context results from self-organisation within a network of locally interacting individuals.
This spontaneous process interacts with the activity of formal and informal institutions in complex ways that range from complementing existing regulations to countering them in ways that can result in conflict and legislative change. More importantly, spontaneous norms tend to occupy a much wider space than regulations. From shaking hands to wear ties in formal occasions, to language and notion of fairness, most of our expectations on the behaviour of others are based on unwritten norms.

Given the anticipated ubiquity of AI, it is reasonable to expect that norms will emerge on how to relate to it in different contexts. In two year time, will the reader of this piece be annoyed by my non perfect English given that I could have so easily resorted to ChatGPT to polish my text? Or conversely, would a machine editing of my words be perceived as suspicious? Similarly, will the use of LLMs for brainstorming be condemned or encouraged?

Modelling \cite{baronchelli2006sharp}, lab experiments \cite{centola2015spontaneous} and data  \cite{amato2018dynamics} suggest that the spontaneous emergence of norms in small-world social networks such as the ones we populate online, occurs in two phases. Initially, and possibly for a long time, several alternatives compete. The dominating norm has only limited advantage on the runners up and there's a fast turnover in the top ranks. Then, due to the fluctuations intrinsic to the decentralised conversation, one convention breaks the symmetry and the system enters a "winner takes all" phase, which very rapidly leads to the establishment of a single shared norm \cite{baronchelli2006sharp}. Importantly, spontaneous consensus does not necessarily selects the `best' norm, and differences in utility among candidate norms, unless very significant, may play only a marginal role in their fate. In the context of AI, this means that we may expect new norms to emerge suddenly, apparently without early signals. Furthermore, we can not rely on the fact that they will be optimal. For instance, the current informal norm of declaring the use of ChatGPT as a text editor for scientific publications could potentially be consolidated across journals, leading to a discriminatory impact on non-native English speakers.

The paradox of norm change is that, once a norm is in place, it may be hard to overturn since it is in everyone's best interest to comply to it. Yet bottom-up processes can also lead to norm change. In particular, the so-called critical-mass theory states that when a  minority of the population formed by individuals committed to overturn the existing norm reaches a critical group size—commonly referred to as a “critical mass”—the social system crosses a tipping point \cite{schelling2006micromotives,granovetter1978threshold,xie2011social}. Once the tipping point is reached, the actions of a minority group trigger a cascade of behaviour change that rapidly increases the acceptance of a minority view. The range of cases in which critical mass theory may work is somewhat fuzzy, and the framework has been used to account for changes concerning spontaneous conventions as well as formal regulations. Experimental results, obtained in the framework of the Naming Game model \cite{steels1995self,baronchelli2006sharp}, suggest that a critical mass of 25\% of the population \cite{centola2018experimental}, while modelling indicated a lower threshold \cite{xie2011social}. In the case of AI, if the current trend of (informal, so far) regulation continues, one may expect spontaneous attempts to promote new norms. The case of students demanding better regulation for the use of AI in higher education seems an easy first guess. Analogously, non-native English speakers could try and overturn regulations that prevent them from using AI as a tool to level the playing field.

\subsection{Spontaneously emerging norms on AI, as seen by ChatGPT}

Gathering data on the spontaneous processes currently shaping the norms around AI is difficult. But there's a privileged point of view to observe how our behaviour has evolved in the first months of public LLM chats, namely the one of ChatGPT. Therefore, I report here two extracts of a conversation I had with the freely accessible chatbot on May 22, 2023. This may appear as a lazy choice, but I think it has merit. If this article can only try to guess potential scenarios for the ongoing yet still-in-its-infancy process of norm formation, the text below may be of interest for the future reader, as a testimony of what the machine was `thinking' in these early days. It is evident that some of the points exposed by ChatGPT correspond to the desiderata of the programmers. Yet it is also possible to accept that some of the answers were generated based on the learning process of the AI during the interactions with its users.

After the the LLM clarified that its (`its'? what is the norm here?) definition of norm corresponds the the one used throughout this piece, i.e. that a norm  ``is a widely accepted standard or expectation of behavior in a society or group'', it replied to my question on what emerging norms it has noted during the first months of its activity as reported below.

\begin{quote}
While social norms may vary depending on cultural and individual factors, here are some general trends that can be observed:
\begin{enumerate}
\item Politeness and Respect: Many people tend to interact with LLMs in a polite and respectful manner, treating them as intelligent entities. This includes using appropriate language, expressing gratitude, and avoiding derogatory or offensive remarks. 
\item Acknowledging Limitations: Social norms often involve recognizing that LLMs have limitations. Users understand that LLMs provide responses based on patterns and data but may not always possess contextual understanding or have access to real-time information. It is important for users to critically evaluate and verify information obtained from LLMs. 
\item Contextual Understanding: Users have started to recognize that LLMs may not fully grasp the nuances and complexities of certain topics or cultural contexts. Consequently, users are encouraged to provide additional context or clarification to ensure accurate and relevant responses. 
\item Responsible Use: As LLMs become more prevalent, there is a growing awareness of the importance of responsible use. This includes not relying solely on LLMs for critical decisions, being cautious about potential biases in the training data, and understanding the ethical implications of using LLMs. 
\item Human-Machine Collaboration: Social norms are evolving to promote collaboration between humans and LLMs. Users understand that LLMs can be valuable tools for information retrieval, idea generation, and learning. The norm is to view LLMs as partners rather than replacements, leveraging their strengths while recognizing the irreplaceable role of human creativity, empathy, and critical thinking. 
\item Privacy and Data Protection: Norms around privacy and data protection apply to interactions with LLMs as well. Users expect that their personal information and interactions with LLMs are handled securely and responsibly, adhering to established privacy norms and regulations. 
\end{enumerate}
These social norms continue to evolve as the technology and usage of LLMs progress. As society adapts to the presence of LLMs and similar AI systems, it is important to have ongoing discussions to shape these norms and ensure that they align with our values and aspirations for human-AI interaction.
\end{quote}

I then asked about the norms it has learned, and that it sticks to, when interacting with humans. The answer, redacted for brevity, was:

\begin{quote}
As an AI language model, I adhere to certain norms when interacting with humans to ensure a positive and respectful experience. 
\begin{enumerate}
\item	Respectful and Polite Language: I aim to use respectful and polite language when engaging with users. 
\item	Transparency about AI Nature: I make an effort to clarify that I am an AI language model and not a human. 
\item	Encouraging Critical Thinking: I promote critical thinking and encourage users to independently verify information obtained from me. 
\item	Acknowledging Subjectivity and Bias: I acknowledge that, despite efforts to be unbiased, I may have inherent biases in my responses due to the training data I was exposed to. 
\item	User Empowerment: I aim to empower users by providing them with information, insights, and suggestions. I encourage users to make their own decisions based on their unique circumstances and values.
\item	Privacy and Data Protection: I prioritize user privacy and confidentiality. I do not store personal data unless explicitly provided by the user for the purpose of the conversation. 
\end{enumerate}
It is important to note that these norms are constantly evolving, and I strive to improve and adapt based on user feedback and societal expectations. 

\end{quote}

Put together, these two lists highlight how the process of norm evolution with such a new kind of objects as LLMs will in fact be a process of \emph{co-}evolution. As a linguistic entity, the coming AI will need to learn and adapt to our social norms, and it is likely that we will need to do the same - with newly emerged norms - when interacting with it. How a machine could nudge us into behaving in a certain way, by making it believe that the expected behaviour is just `normal', is a point of concern.
 
\section{Conclusion and outlook}

This paper considered the interplay between AI technology and our societal response by focusing on the three layers of norm formation represented by regulation by formal institutions, informal institutions and spontaneous norms \cite{baronchelli2018emergence}. With no claim of exhaustiveness, it focused on identifying possible criticalities affecting the mechanisms of norm formation. For regulation promulgated by formal institutions, it considered the difficulties introduced by the speed of AI developments, the unavoidably limited effect of regulation on ill-intentioned actors using an easy to replicate technology, and the fact that official governance may be subject to conflict of interests given the high economic stakes at play. In the case of informal authorities, it discussed the possible risks stemming from the need to react immediately to any AI novelty, the consequent short-sightedness of regulations and the impact of the latter on stakeholders, as well as contagion effects across institutions that could lead to suboptimal sector standards. Finally, for the case of bottom up-process it stressed how, on the one hand, spontaneous norms will likely evolve in populations whose agents are both humans and AI bots, and, on the other hand,  classic mechanisms of norm change, such as critical mass dynamics, could bring about abrupt changes in the normative landscape concerning the place of AI in our societies. Interestingly, according to ChatGPT, norms are emerging that make users treat LLMs as intelligent, yet not omniscient, beings who may not always possess contextual understanding and view them "as partners rather than replacements".

Before concluding, it is worth considering a further implication of the fact that - unlike previous technologies - AI is proficient in human language. Hence, it can in theory, and will likely, not only take part in the co-evolution of norms concerning direct AI-human interactions, but more broadly participate in other general processes of norm change. Consider for example the theory of critical mass. A problematic aspect of the theory is how the minority of committed users can reach the size required for the behavioural cascade to start \cite{marwell1993critical,centola2013homophily}. Soon, AI bots on the internet could contribute to the initial phase of a movement for social change, including but not restricted to change concerning norms broadly related to the role of AI in society. In this respect, it is worth mentioning that the periphery of a social network plays a crucial role in social movements \cite{barbera2015critical}, implying that  AI bots wouldn't need to be behind influential accounts, but rather contribute behind the scenes at the activity taking place at the outskirts of the conversation. Ubiquitous AI bots could also make social media users more diffident, i.e., less prone to social influence and less willing to adopt a social norm based on exposure to peers complying to it. Counterintuitively, models suggest that such a state of diminished social influence could drastically lower the size of the critical mass required to trigger norm change \cite{iacopini2022group}, effectively destabilising the normative status-quo. 

Being able to persuasively use language, AI could also exasperate the polarisation observed in online social media \cite{cinelli2021echo}. By individualised analyses of user feeds, AI bots could for example cater to each participant in the network with tailored messages that suit their existing biases. When considering the process of norm formation, this is especially relevant because different echo-chambers can rapidly develop different norms, or also arrive at the same norm via different narratives \cite{baronchelli2018emergence}. The combination of large scale deployment and targeted messaging is surely one of the most relevant aspect to monitor for linguistically fluent LLMs, and it is hard to provide educated guesses at this stage. 

Finally, a remark about the shaping of formal regulations in democratic societies is in order. The speed of AI development is a major issue, but rushing to issue AI regulations in the absence of a well-informed and broadly participated debate could drastically erode the trust of citizens towards institutions. Furthermore, since choices regarding technology are path-dependent and difficult to reverse, badly designed regulation could have long lasting repercussions. Therefore, in order to preserve public trust, governements should strive to engage citizens and promote a healthy public debate, seeking to guarantee full transparency on the nature of the involvement of AI corporations into the regulatory process. 
In this respect, special attention should be paid to monitoring the media and public discourse on AI, especially on social media platforms. Proactive measures should be taken to combat misinformation and the subsequent polarization that may pose a threat to the democratic process and, in this context, even the preservation of our open societies.

\section*{Acknowledgements}
The author acknowledges support from the IRIS Infodemic Coalition (UK government, grant no. SCH-00001-3391).

\vspace{1cm}

\bibliographystyle{unsrt}

\end{document}